\documentclass[traditabstract]{aa}
\usepackage{subfigure}
\usepackage{longtable}
\usepackage{graphicx}
\usepackage{graphics}
\usepackage{keyval}
\usepackage{trig}
\usepackage{psfig,epsf}
\usepackage{txfonts}
\def\beq{\begin{equation}}
\def\eeq{\end{equation}}

\begin{document}

\title{HALOGAS: Extraplanar gas in NGC 3198}

\author{G. Gentile\inst{1,2} \and G. I. G. J\'ozsa\inst{3,4} \and
  P. Serra\inst{3} \and G. H. Heald\inst{3} \and W. J. G. de
  Blok\inst{3,5} \and F. Fraternali\inst{6,7} \and
  M. T. Patterson\inst{8} \and R. A. M. Walterbos\inst{8} \and
  T. Oosterloo\inst{3,7} }  

\institute{Sterrenkundig Observatorium, Universiteit Gent, Krijgslaan 281, B-9000 Gent, Belgium\\
              \email{gianfranco.gentile@ugent.be}
    \and
            Department of Physics and Astrophysics, Vrije Universiteit
            Brussel, Pleinlaan 2, 1050 Brussels, Belgium
    \and
      Netherlands Institute for Radio Astronomy (ASTRON), Postbus 2, NL-7990 AA Dwingeloo, The Netherlands 
    \and
     Argelander-Institut f\"ur Astronomie, Auf dem H\"ugel 71, D-53121
     Bonn, Germany 
    \and
     Astrophysics, Cosmology and Gravity Centre (ACGC), Astronomy
     Department, University of Cape Town, Private Bag X3, 7700
     Rondebosch, Republic of South Africa 
     \and
     Astronomy Department, University of Bologna, Bologna, Italy
     \and
      Kapteyn Astronomical Institute, University of Groningen, AD
      Groningen, the Netherlands 
     \and
     Department of Astronomy, New Mexico State University, PO Box
     30001, MSC 4500, Las Cruces, NM 88003, USA 
     }

\abstract
{We present the analysis of new, deep H{\sc i}\ observations of the spiral
  galaxy NGC 3198, as part of the HALOGAS (Westerbork Hydrogen
  Accretion in LOcal GAlaxieS) survey, with the main aim of
  investigating the presence, amount, morphology and kinematics of
  extraplanar gas. We present models of the H{\sc i} observations of NGC
  3198: the model that matches best the observed data cube features a
  thick disk with a scale height of $\sim$3 kpc and an H{\sc i} mass of about
  15\% of the total H{\sc i} mass; this thick disk also has a decrease in
 rotation velocity as a function of height (lag) of 7--15 km s$^{-1}$
 kpc$^{-1}$ (though with large uncertainties). This extraplanar gas is
 detected for the first time in NGC 3198. Radially, this
 gas appears to extend slightly beyond the actively
 star-forming body of the galaxy (as traced by the H$\alpha$ emission),
 but it is not more radially extended than the outer, fainter
 parts of the stellar disk. Compared to previous studies, thanks
 to the improved sensitivity we trace the rotation curve out to
 larger radii. We model the rotation curve in the framework of
 MOND (Modified Newtonian Dynamics) and we confirm that, with the
 allowed distance range we assumed, fit quality is modest in this
 galaxy, but the new outer parts are explained in a satisfactory
 way. 
}


\maketitle

\section{Introduction}
\label{sect_intro}

The last couple of decades have seen a wealth of observational data
revealing the presence of material outside the plane of disk galaxies
(see Sancisi et al. 2008 and Putman et al. 2012 for a
review), e.g. hot X-ray emitting gas (e.g. T\"ullmann et al. 2006, Li et
al. 2008), ionised gas (Collins \& Rand 2001, Rossa \& Dettmar 2003),
dust (Howk \& Savage 1999, M\'enard et al. 2010), and neutral hydrogen
(Fraternali et al. 2002; Oosterloo et al. 2007; Heald et
al. 2011). This material is found to be often (but not always, in the
case of the H{\sc i}) closely associated to star formation in the galaxy
disk.

\begin{figure*}
\begin{center}
\includegraphics[width=0.99\textwidth]{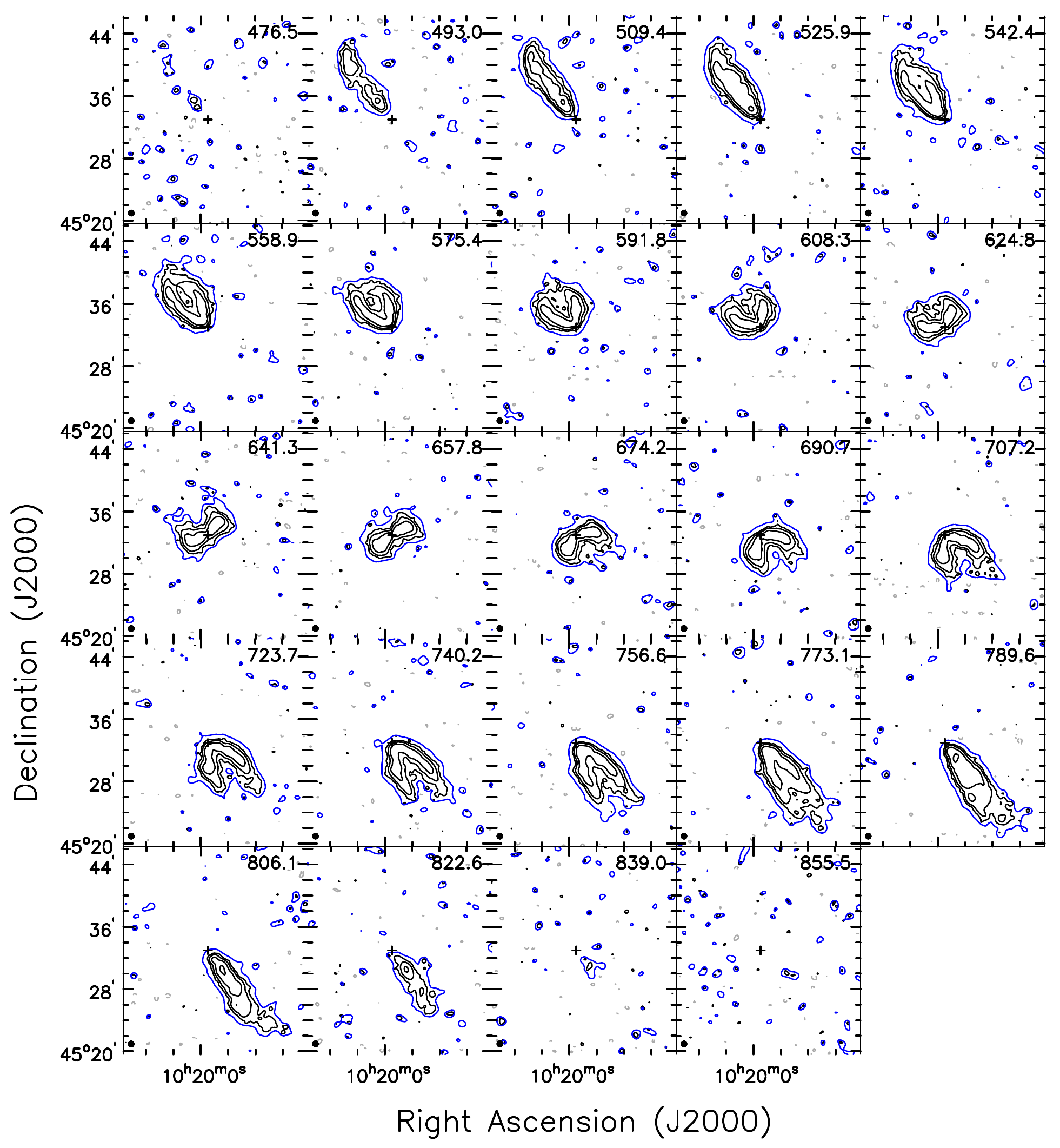}
\end{center}
\caption{
Channels maps of the H{\sc i} observations of NGC 3198 presented here. The
velocity of each channel map is indicated in the top right corner of
each panel (every fourth channel map is shown). The beam is indicated
in the bottom left (35.2 $\times$ 33.5 arcsec). The contours are -0.42, 0.42
(2$\sigma$), 2.1, 10.5, 52.5 mJy beam$^{-1}$. The blue contour is the 0.42 mJy
beam$^{-1}$ (2$\sigma$) level of the 60 arcsec resolution cube. Negative contours
are dashed grey. The cross indicates the galaxy centre. 
}
\label{channels}
\end{figure*}

The galactic fountain is an example of a framework that attempts to
explain the presence and properties of extraplanar gas: gas is
expelled to large distances from the disk by supernova explosions and
adiabatic expansion, and then it falls back onto the galaxy disk due
to radiative cooling (Shapiro \& Field, 1976; Bregman, 1980). For the
conservation of angular momentum, there must be a decrease of the
rotation velocity (also known as ``lag'') with distance from the
plane. However, ballistic models of the interplay between the
thin disk and the extraplanar gas
(e.g. Fraternali \& Binney 2006) show that there must be additional
mechanisms governing the kinematics of extraplanar gas, because these
simple models obtain lags that are too shallow compared with
observations. A possibility is the interaction between the uplifted
gas and a corona of slowly-rotating hot gas, which would explain the
magnitude of the observed lags (Marinacci et al. 2011). Other
explanations for the properties of extraplanar gas 
involve e.g. thermal instabilities in the corona (Kaufmann et al. 2006, but see
Binney et al. 2009), pressure gradients or the effect of the magnetic
field (Benjamin 2002).

Observational evidence for extraplanar lagging H{\sc i} gas comes from both
edge-on and moderately inclined galaxies. Edge-on galaxies allow us to
assess the extent of this gas and the magnitude of the lag,
whereas in moderately inclined galaxies one can investigate the local
connection between star formation
and extraplanar gas, and one can study in detail its kinematics. 
Observationally, in moderately inclined galaxies the
very existence of the lag allows the disentangling of extraplanar gas
from gas that resides in the plane, e.g. in NGC 2403 (Fraternali et
al. 2002), NGC 4559 (Barbieri et al. 2005), and NGC 6946 (Boomsma et
al. 2008). 

In Heald et al. (2011) we presented the HALOGAS (Westerbork Hydrogen
Accretion in LOcal GAlaxieS) survey, a systematic investigation of
extraplanar gas in 22 spiral galaxies, using very deep H{\sc i}
observations. The main goal of HALOGAS is to investigate the amount
and properties of extraplanar gas. The target galaxies were selected
to be nearby and to fall in one of two categories: edge-on galaxies
(inclination $i\geq 80^{\circ}$) and moderately-inclined galaxies
(50$^{\circ} \leq i \leq 75^{\circ}$). 

NGC 3198 is one of the moderately-inclined galaxies in the HALOGAS
sample. It is a late-type spiral that is traditionally considered as
one of the benchmark galaxies for rotation curve studies, because its
H{\sc i} kinematics is regular and symmetric, its H{\sc i} disk is very extended
and its inclination angle (around 70$^{\circ}$) is ideal for deriving the
rotation curve. Many studies have focussed partly or completely on
the kinematics of NGC 3198, e.g. van Albada et al. (1985), Begeman
(1989), Blais-Ouellette et al. (2001), Bottema et al. (2002), de Blok
et al. (2008), Sellwood \& S\'anchez (2010), Gentile et al. (2011). The
distance of NGC 3198 has been determined to be 13.8 Mpc using Cepheids
(Kelson et al. 1999, Freedman et al. 2001, see also Macri et
al. 2001), usually considered to be one of the most precise distance
indicators. 

In the present paper we aim to characterise the presence, amount, and
kinematics of any extraplanar gas in NGC 3198, using the most
sensitive H{\sc i} data available and a careful tilted-ring model based
analysis. We also look for a link between the presence of this
extraplanar gas and star formation in this galaxy. Additionally, we
present detailed kinematic modelling of the observations, as well as
a mass decomposition of the rotation curve.

\section{Data acquisition and reduction}

\subsection{H{\sc i} data}

We summarise here how the data were acquired and reduced, and we refer
the reader to Heald et al. (2011) for a more thorough description of
the observations and data reduction, of which we only
describe the main points. Information about NGC 3198 and the
observations presented here is given in Table \ref{tab-obs}.

\begin{table}
\caption{The galaxy NGC 3198 and the observations presented here. The
  beam FWHM refers to the cube modelled in this paper. References are: 1:
  HyperLeda; 2: Kelson et al. (1999), Freedman
et al. (2001); 3: this work.}              

\label{tab-obs}      
\centering                                      
\begin{tabular}{l l l}  
\hline\hline     
Parameter                & Value & Reference \\    
\hline
Hubbe type & Sc & 1  \\
Adopted distance & $13.8\pm 1.5 $ Mpc  & 2 \\
B-band magnitude & $10.90 \pm 0.07$  & 1 \\
\hline                       
Synthesised beam FWHM & 35.2 $\times$ 33.5 arcsec& 3\\
mJy/beam to Kelvin & 0.51 &  3 \\ 
On-source time & 132 hr & 3\\
Velocity resolution & 4.12 km s$^{-1}$&  3\\
RMS noise & 0.21 mJy beam$^{-1}$ &3\\
\hline       
\end{tabular}
\end{table}

The observations of neutral hydrogen (H{\sc i}) were obtained at the WSRT
(Westerbork Synthesis Radio Telescope) in the ``maxi-short''
configuration, for 10 $\times$ 12 hours with a bandwidth of 10 MHz divided
into 1024 channels, and two linear polarizations. The data were
reduced using the software package Miriad (Sault et al. 1995). After
flagging, calibration and continuum subtraction, the data were Fourier
inverted to obtain a data cube. When inverting, a Gaussian taper with
an image-plane width of 30 arcseconds was applied to the uv-data,
which were also Hanning-smoothed to a velocity resolution of 4.12 km
s$^{-1}$. The data cubes obtained in this manner have an rms noise of 0.21
mJy beam$^{-1}$. Avoiding to apply the taper gives a higher-resolution cube
(18.9 $\times$ 13.6 arcsec) but this cube was not used for modelling as one
of the main goals of this paper is to investigate the faint diffuse
emission, whose signal-to-noise ratio is higher when applying the
taper and thus somewhat degrading the resolution. 

The CLEAN deconvolution of the dirty cube was done in two steps:
first, a CLEAN mask was defined (for every channel map) where
presence of real emission can be established with confidence. In these
regions the CLEAN deconvolution was performed down to approximately
1$\sigma$. Then, the channel maps were CLEANed on the whole field, down to
about 2$\sigma$. Finally, the channel maps were restored using a Gaussian
beam with a FHWM size of 35.2 $\times$ 33.5 arcsec.

\begin{figure}
\begin{center}
\includegraphics[width=0.47\textwidth]{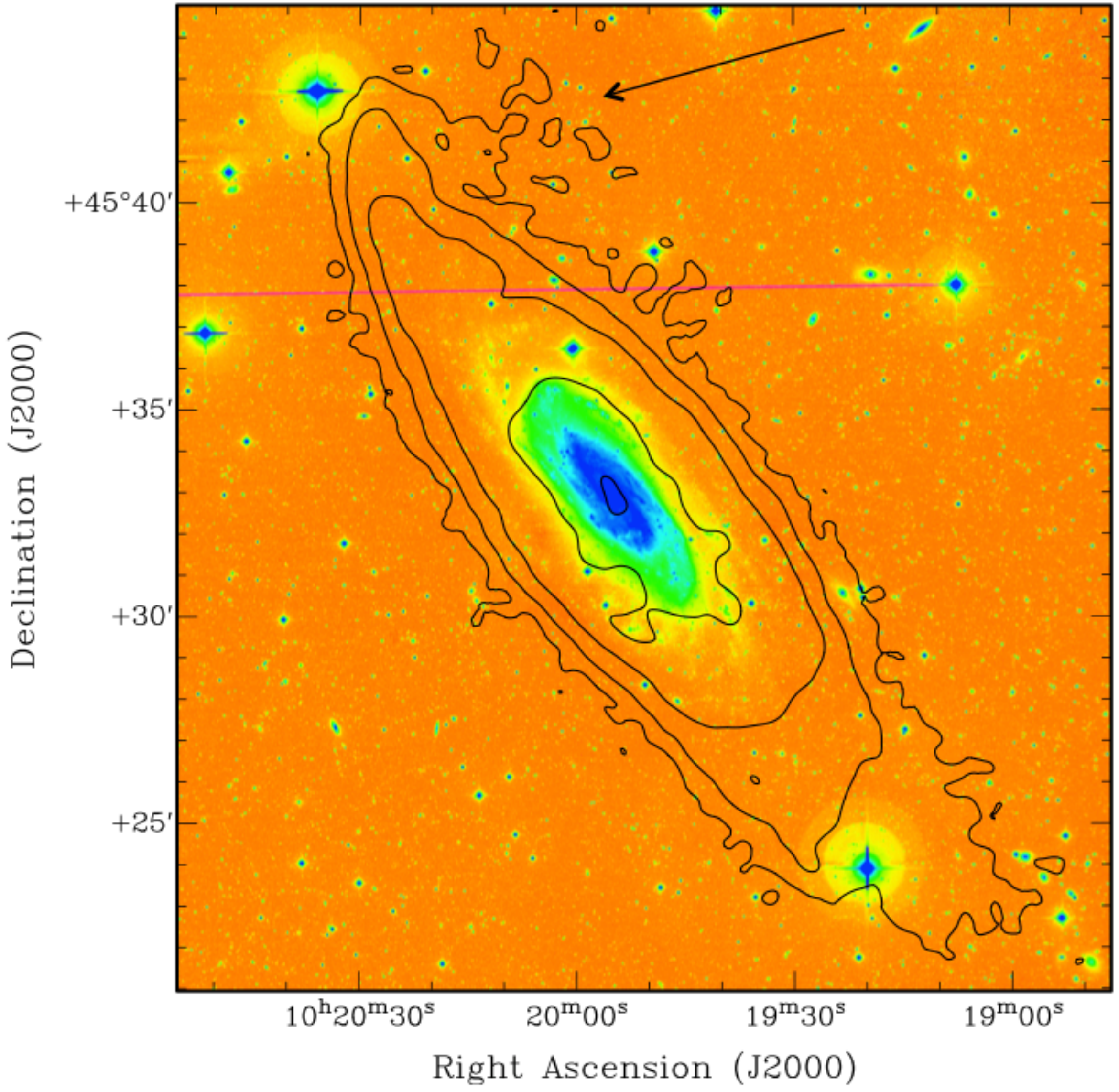}
\end{center}
\caption{
Total H{\sc i} map (contours) superimposed onto a false colour r'- band
HALOSTARS image (see text). Contours are 0.1, 1, 5, and 15 $\times$10$^{20}$ atoms
cm$^{-1}$. The black arrow indicates the tentatively detected H{\sc i} described
in Section \ref{sect_observed}.} 
\label{n3198_mom0_halostars}
\end{figure}

\begin{figure*}
\begin{center}
\includegraphics[width=0.95\textwidth]{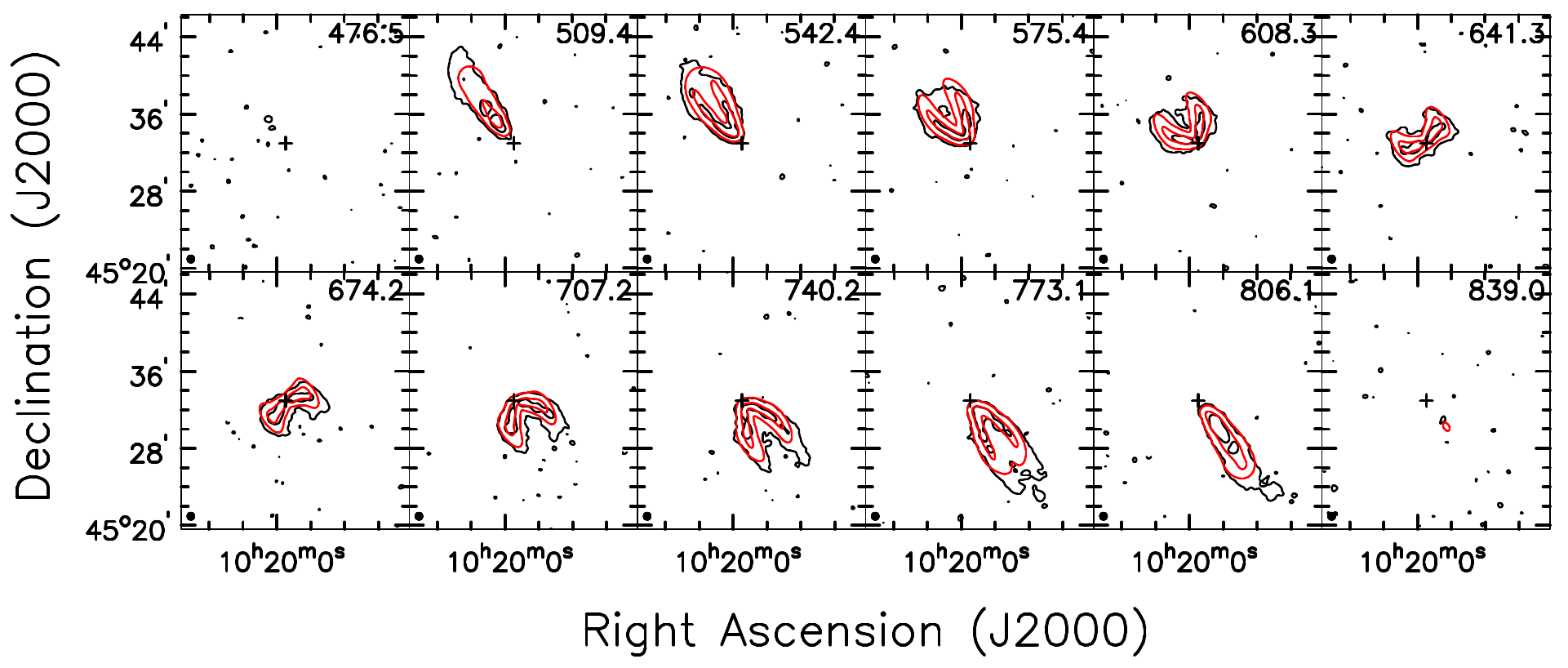}
\end{center}
\caption{
Channels maps of NGC 3198: observed (black) and modelled (red), using
the rotation curve and inclination parameters from de Blok et
al. (2008). Symbols are the same as Fig. \ref{channels}. For better
readability we have plotted fewer contours: -0.5, 0.5 (2.5$\sigma$),
and 20 mJy beam$^{-1}$. Also, compared to Fig. \ref{channels}, fewer
channels are shown.
}
\label{channels_data_things}
\end{figure*}

\begin{figure*}
\centering
\begin{minipage}[b]{0.32\textwidth}
\includegraphics[width=\textwidth]{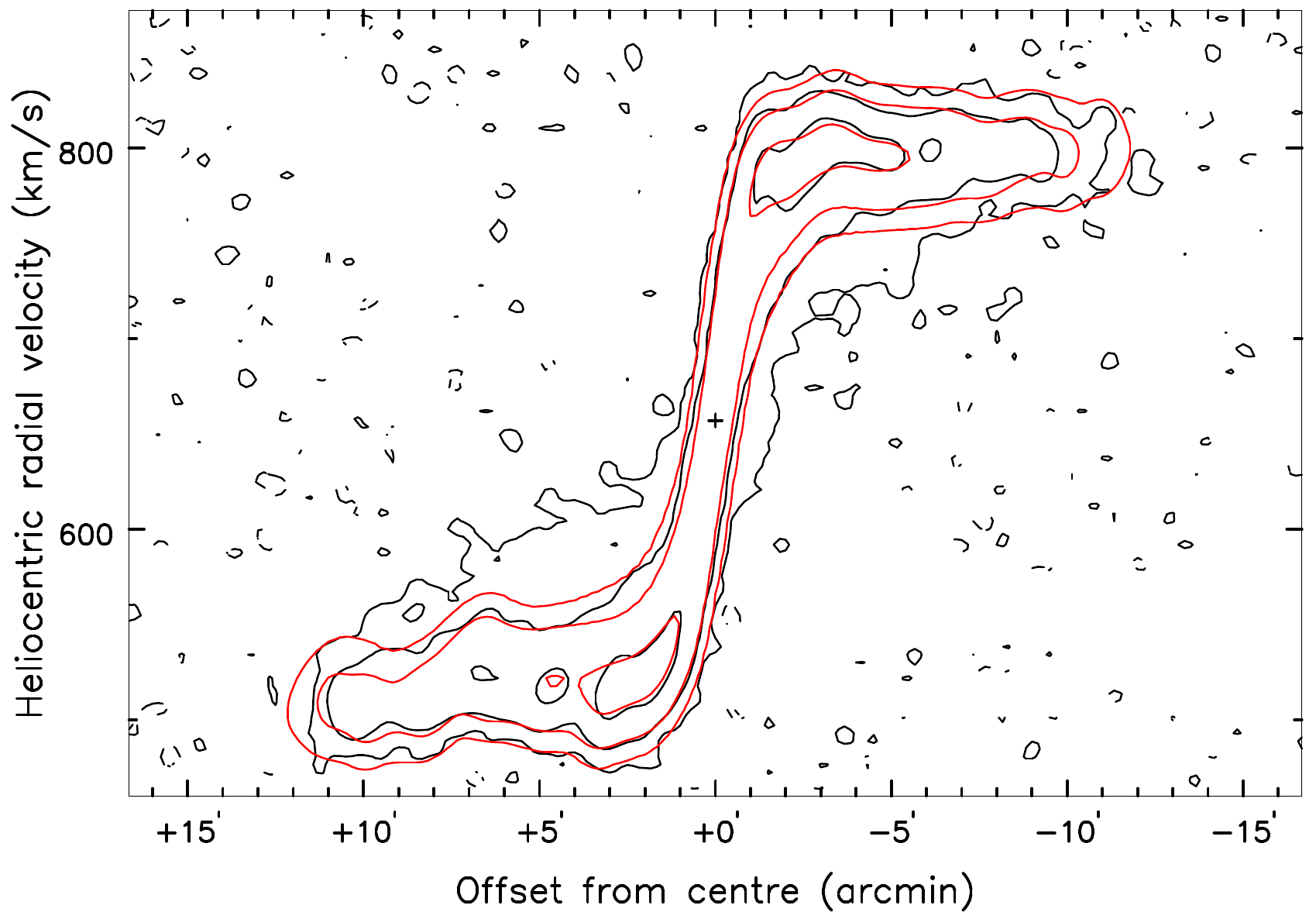}
\end{minipage}
\hspace{0.01em}
\begin{minipage}[b]{0.32\textwidth}
\includegraphics[width=\textwidth]{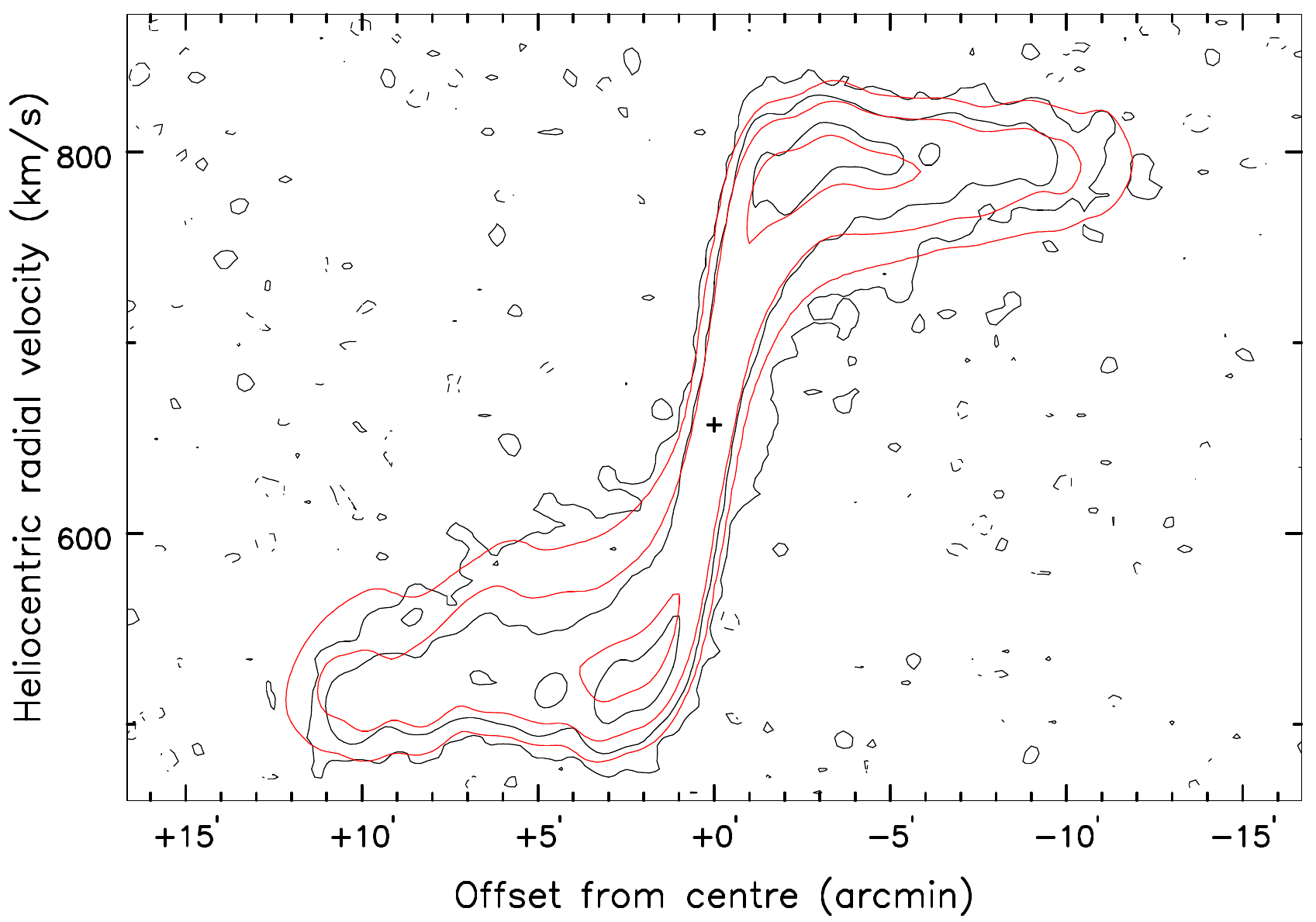}
\end{minipage}
\hspace{0.01em}
\begin{minipage}[b]{0.32\textwidth}
\includegraphics[width=\textwidth]{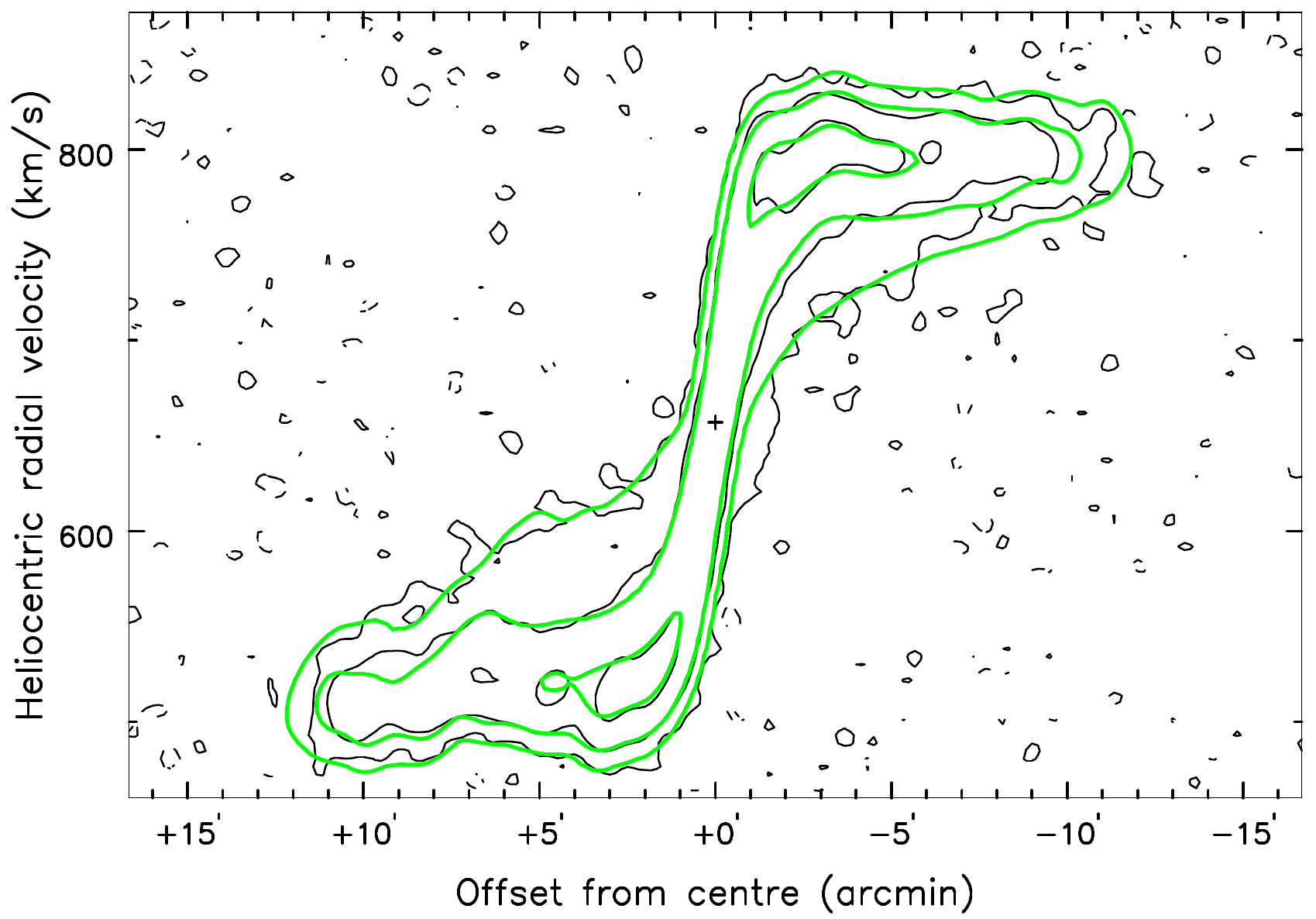}
\end{minipage}
\caption{Position-velocity diagrams of NGC 3198 along the major axis,
  considering a position angle of 213$^{\circ}$ : observed (black) and modelled
  (left and middle panel: red; right panel: green), derived using the
  steps outlined in section \ref{sect_models}. Left: without lagging extraplanar
  gas. Middle: with a single 1-kpc thick disk with a lag. Right: with
  a separate disk of lagging extraplanar gas. Contours are -0.3, 0.3 ($\sim$1.5$\sigma$),
3, and 
30 mJy beam$^{-1}$. The cross indicates the galaxy
  centre.}
\label{pvds}
\end{figure*}

\begin{figure*}
\begin{center}
\includegraphics[width=0.97\textwidth]{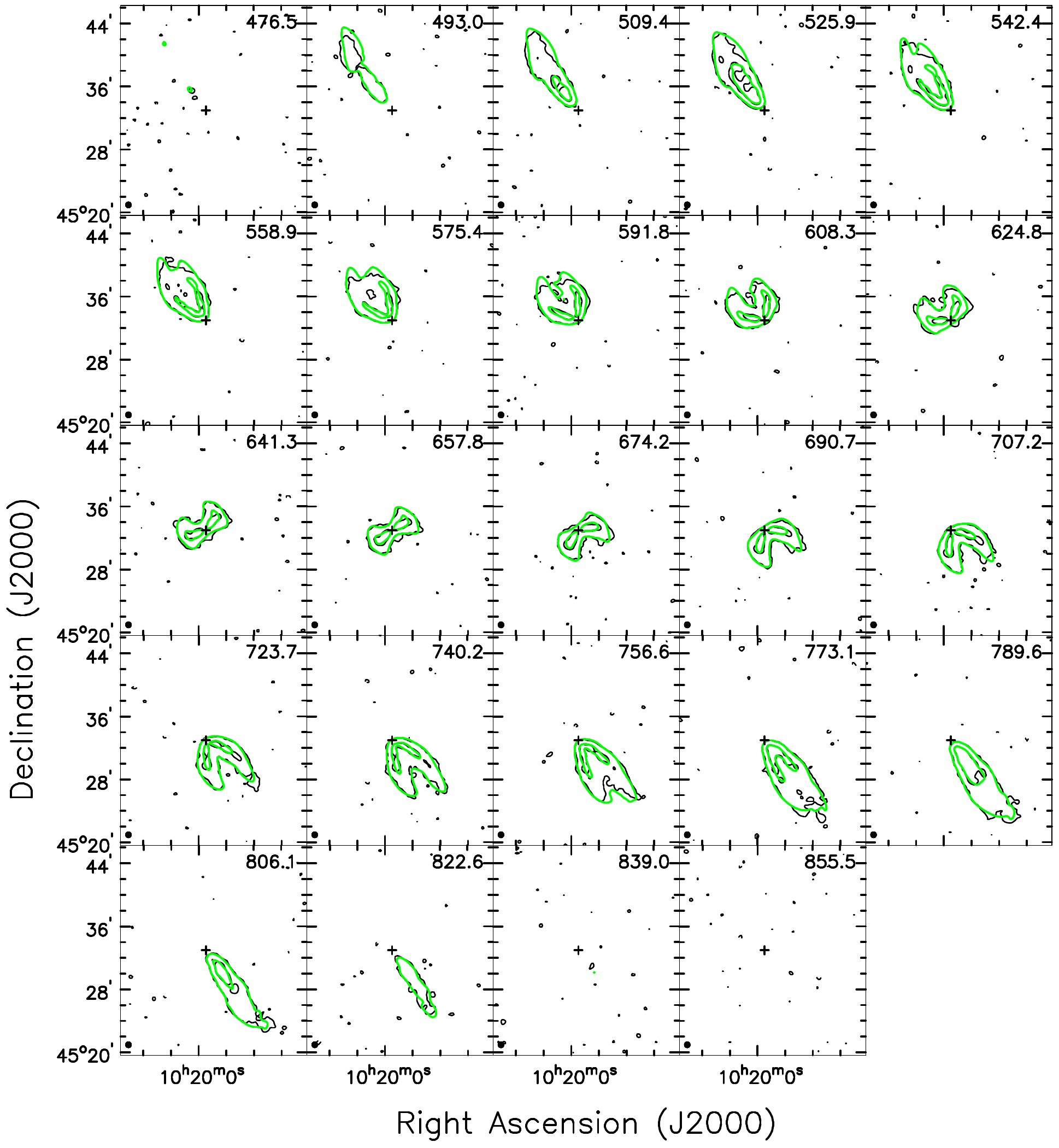}
\end{center}
\caption{
Channels maps of NGC 3198: observed (black) and modelled (green), using our final model parameters. Symbols are the same as Fig. \ref{channels}. For better readability we have plotted fewer contours: -0.5, 0.5 (2.5$\sigma$), and 20 mJy beam$^{-1}$. 
}
\label{channels_data_final}
\end{figure*}

\subsection{Optical photometry}

We obtained photometric observations of NGC 3198 within the framework
of HALOSTARS, which is a deep
optical survey of the HALOGAS targets with the Isaac Newton Telescope
(INT). NGC 3198 was observed in the
r'-band on February 4, 2010, and the total on-source
time amount was 3600 s. A coadded image was obtained using the THELI
\footnote{http://www.astro.uni-bonn.de/~theli/} imaging reduction 
pipeline (Schirmer et al. 2003, Erben et al. 2005). After the
overscan- and bias correction, the images were
flat-fielded, super-flat-fielded and defringed. We also masked non-Gaussian
noise-features such as cosmic rays. Using SCAMP (Bertin 2006), the
resulting images were
photometrically calibrated, background-subtracted and
co-added, taking the sky background variation in the
individual chips into account and using the 2MASS catalogue to
solve for astrometric distortion and for the
relative sensitivity of the chips. Using the SDSS (DR8) catalogue, 
we determined the zero point to be 24.67
$\pm$ 0.02 mag. From an rms noise of 
0.023 ADU/s we thus find a 1-$\sigma$ level of 26.4 mag over 1 square
arcsec. To enhance faint extended features, the image 
was convolved with a Gaussian with a FWHM 3 arcsec. 

\subsection{H$\alpha$ imaging}
\label{sect_halpha}

The H$\alpha$ observations of NGC 3198 were made at the Kitt
Peak National Observatory (KPNO) as part of an observing run aimed at
obtaining wide-field optical images of the HALOGAS sample, 
to investigate the relation between extraplanar gas and star
formation. We observed
over 4 nights in January 2012 using the Mosaic 1.1 instrument on the
4-meter telescope. We used an H$\alpha$ filter centered at 6574.74
\AA  ~and a  FWHM of 80.62 \AA ngstroms, which includes [NII] lines at 6548
and 6584 \AA.  We exposed in the 5 dither pattern with six minutes per
exposure for a total of 30 minutes.  R-band images (10 minutes total)
were also acquired for continuum subtraction.


\section{Observed H{\sc i} data}
\label{sect_observed}

The data cube is shown in Fig. \ref{channels}, for the lowest contour we also
plotted the low-resolution (60 arcsec) cube. The channel maps show a
regularly rotating disk and some asymmetries (e.g., the south-west
side is slightly more extended).

The total H{\sc i} map was obtained by computing the 0th moment of a masked
version of the data cube, which was determined by masking out spurious
emission (i.e., emission that is not above 2$\sigma$ in at least three
consecutive channels of the 60 arcsec resolution
cube). Fig. \ref{n3198_mom0_halostars} shows 
the total H{\sc i} map superimposed with a HALOSTARS image.  

From the primary beam corrected data cube we find a total H{\sc i} flux
of 239.9 Jy km s$^{-1}$, which translates into a total H{\sc i} mass of 1.08
$\times$ 10$^{10}$ M$_{\odot}$
assuming a distance of 13.8 Mpc. This H{\sc i} mass is slightly (6\%) higher
than the value given in Walter et al. (2008) from the THINGS 
(The H{\sc i} Nearby Galaxy Survey) survey, which might partly be
due to amplitude calibration uncertainties, and partly to the fact
that we detect some extraplanar gas.

In Fig. \ref{n3198_mom0_halostars} one can see that our data are more
sensitive to faint extended emission than previous observations. At a
resolution of 30 arcsec, we find meaningful emission down to $\sim$1
$\times$ 10$^{19}$ atoms cm$^{-2}$. Despite their different angular resolutions, this can
be compared with the lowest contour given by Begeman (1989) (0.5 $\times$
10$^{20}$ atoms cm$^{-2}$) and de Blok et al. (2008) (1 $\times$
10$^{20}$ atoms cm$^{-2}$). Note 
that the latter study focussed more on having a high spatial
resolution than on being sensitive to extended emission. Comparing our
map with the one in Begeman (1989), we now confirm the existence of
the south-west extension hinted 
at by Begeman's 0.5 $\times$ 10$^{20}$ atoms cm$^{-2}$ contour (see his
Fig. 4). However, the total H{\sc i} map presented here extends further out,
especially in the south-west side of the galaxy, and we tentatively
detect some very faint emission to the north of the north-eastern side
(visible also in Fig. \ref{channels}, channels at 493.0 km s$^{-1}$ to
558.9 km s$^{-1}$). We estimate the mass of this H{\sc i} structure
(indicated with a black 
arrow in Fig. \ref{n3198_mom0_halostars}) to be about 5 $\times$ 10$^6$ M$_{\odot}$. We
note that this is the only 
feature that clearly does not belong to the thin or thick disk. In
many galaxies there is evidence for more of such features, indicating
some amount of accreting H{\sc i}. 
  
At a projected distance of about 120
kpc from NGC 3198 we also detect in H{\sc i} a galaxy pair, identified in NED (NASA/IPAC
Extragalactic Database) as VV 834. One of its two components has a radial velocity from
the Sloan Digital Sky Survey of 566 $\pm$ 256 km s$^{-1}$. We find an H{\sc i} mass of 2.1 $\times$
10$^{7}$ M$_{\odot}$ for the galaxy pair. The centres of the two galaxies are only $\sim$ 19
arcsec apart, and the H{\sc i} emission is barely resolved, with a total extent of about 1.5--2
arcmin and spreading over 80 km s$^{-1}$, from $\sim 540$ km s$^{-1}$
to $\sim 620$ km s$^{-1}$.

\begin{figure}
\begin{center}
\includegraphics[width=0.45\textwidth]{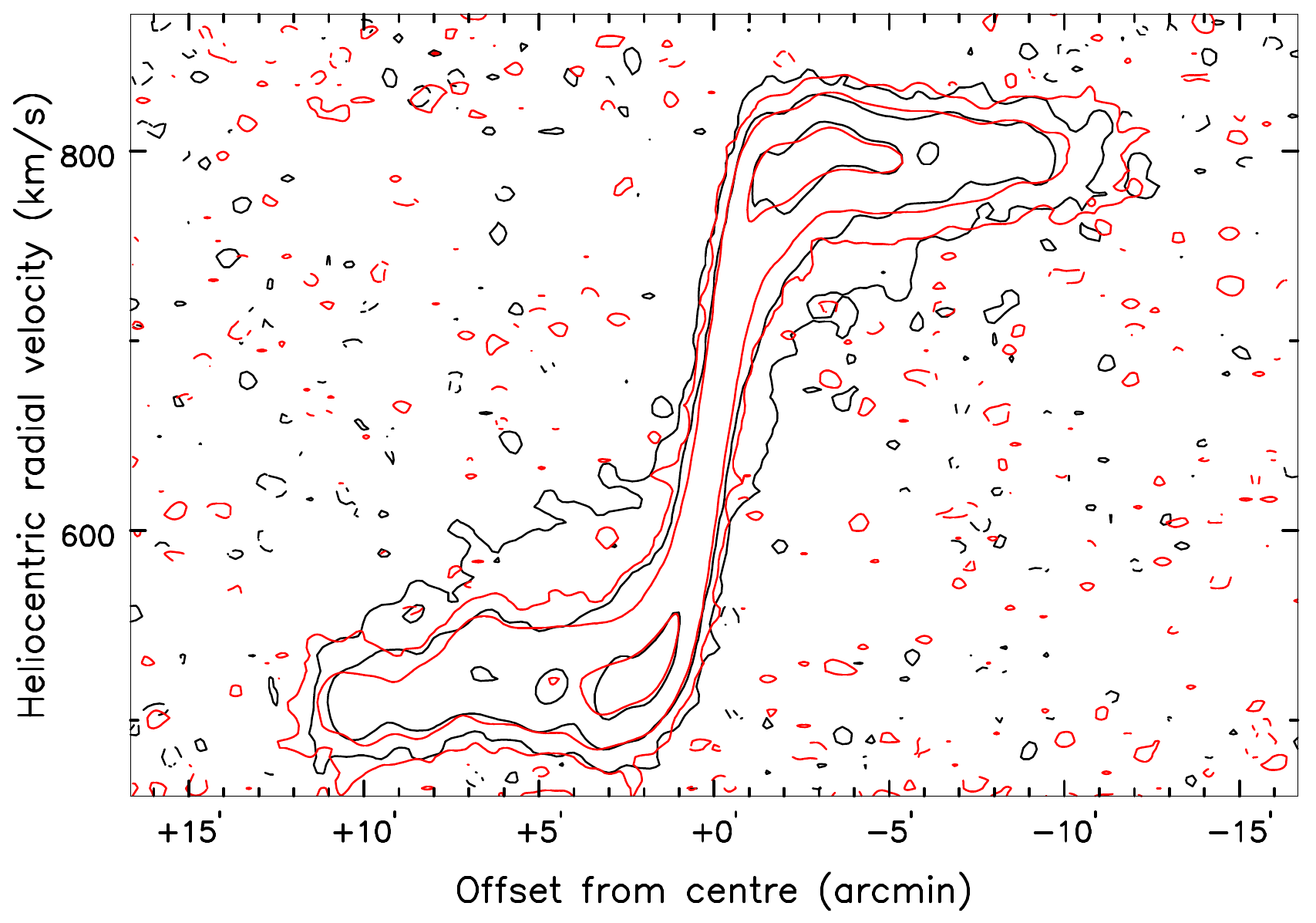}
\end{center}
\caption{
Position-velocity diagram (pvd) of NGC 3198 along the major axis,
considering a position angle of 213$^{\circ}$. The observed pvd is
shown in black and the model is in red. The model shown here is
derived with a thin disk only (plus noise), convolved with the 
dirty beam, CLEANed and restored as described at the end of section
\ref{sect_models}. Contours are -0.3, 0.3 ($\sim$1.5$\sigma$), 3, and
30 mJy beam$^{-1}$. 
}
\label{pvd_dirtybeam}
\end{figure}

\begin{figure}
\begin{center}
\includegraphics[width=0.49\textwidth]{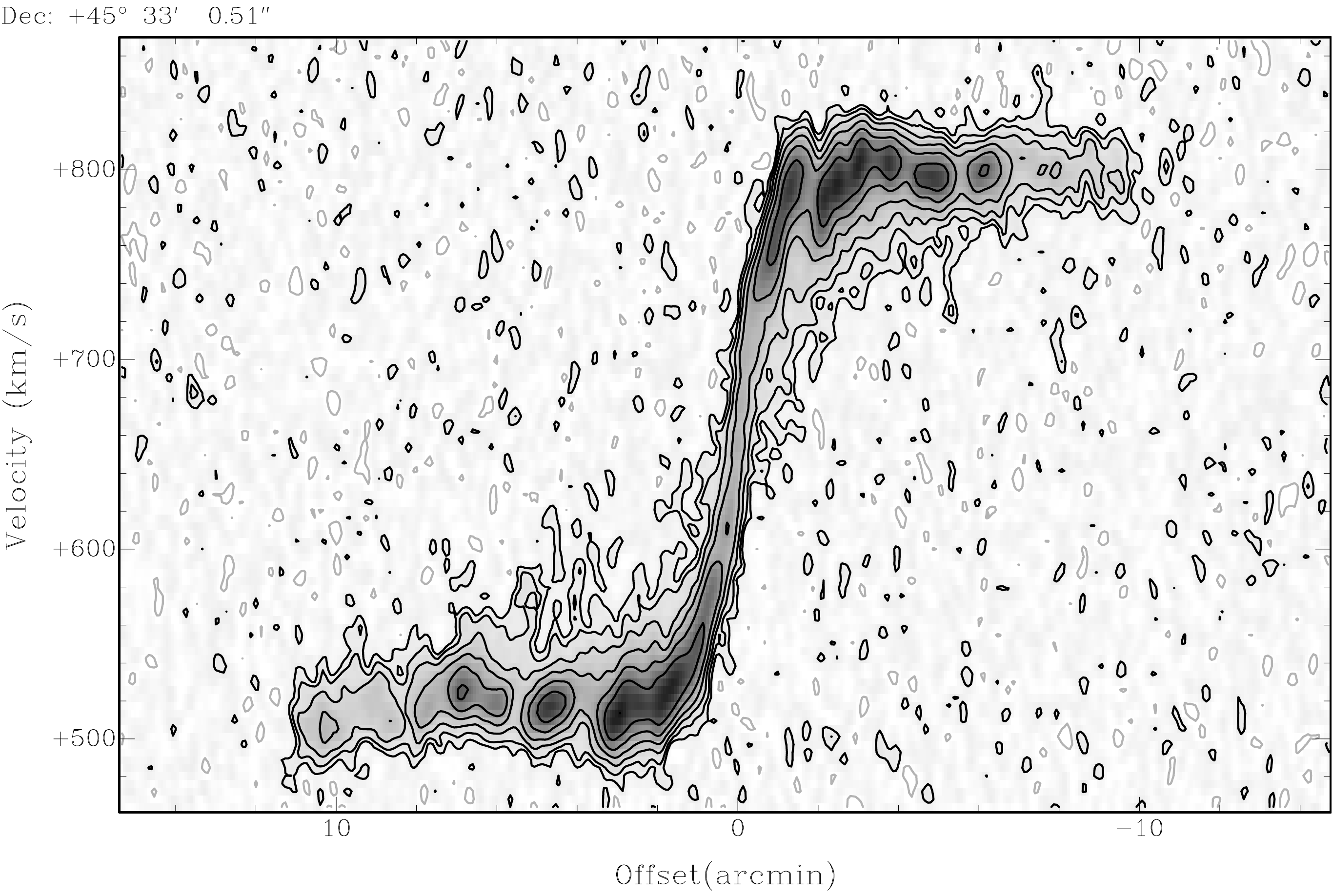}
\end{center}
\caption{
Position-velocity diagram (pvd) of the high-resolution cube (18.9
$\times$ 13.6 arcsec) of NGC 3198 along the major axis,
considering a position angle of 213$^{\circ}$. Contours are -0.29, 0.29
($\sim$1.8$\sigma$), 0.58, 1.16, ... mJy beam$^{-1}$. 
}
\label{pv_majax_hires}
\end{figure}

\begin{figure}
\begin{center}
\includegraphics[width=0.47\textwidth]{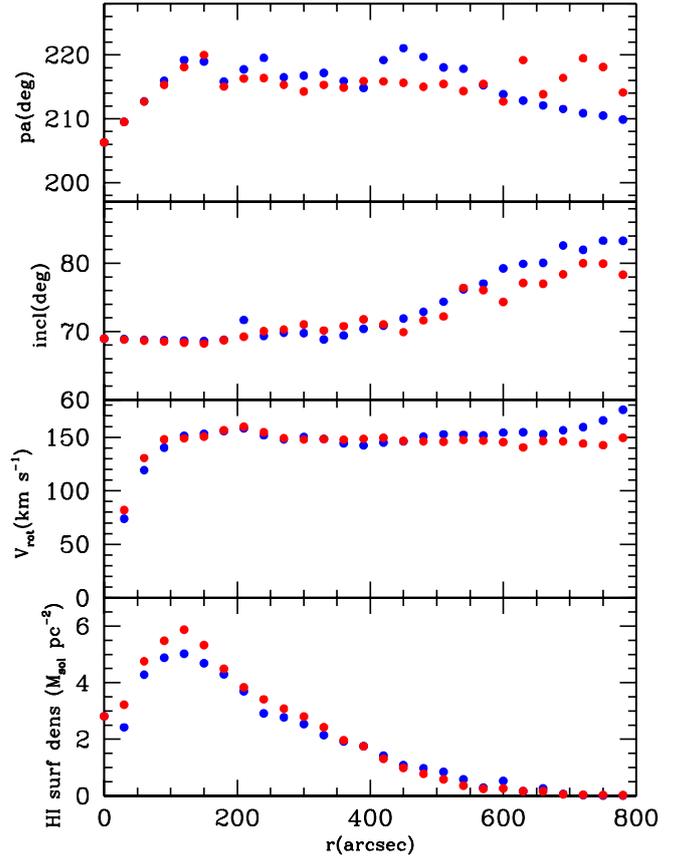}
\end{center}
\caption{
Radial dependence of the parameters of the best-fit model data cube
presented in this paper. Blue represents the approaching (north-east)
side and red is the receding (south-west) side. ``pa'' is the position
angle, ``incl'' is the inclination, ``V$_{\rm rot}$'' is the rotation
velocity, and ``H{\sc i} surf dens'' is the H{\sc i} surface density.
}
\label{best_model_param}
\end{figure}

\begin{figure}
\begin{center}
\includegraphics[width=0.47\textwidth]{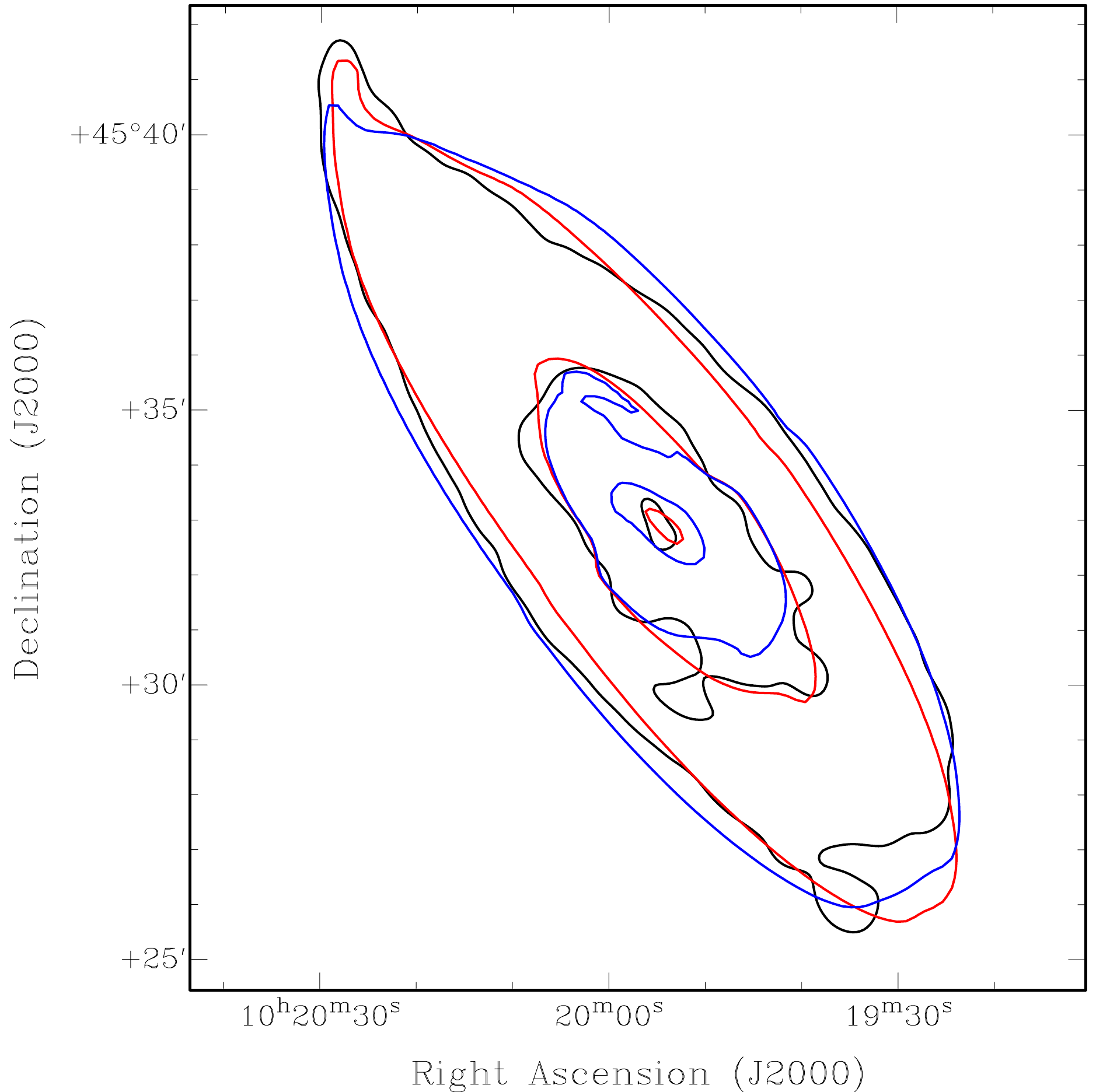}
\end{center}
\caption{
Comparison total H{\sc i} maps: observed (black), with all the inclination
of our best model plus two degrees (red) and minus two degrees (blue). Contours are
3$\times$10$^{20}$ and 1.5$\times$10$^{21}$ atoms cm$^{-2}$.
}
\label{mom0_comparison_incl}
\end{figure}

\begin{figure}
\begin{center}
\includegraphics[width=0.49\textwidth]{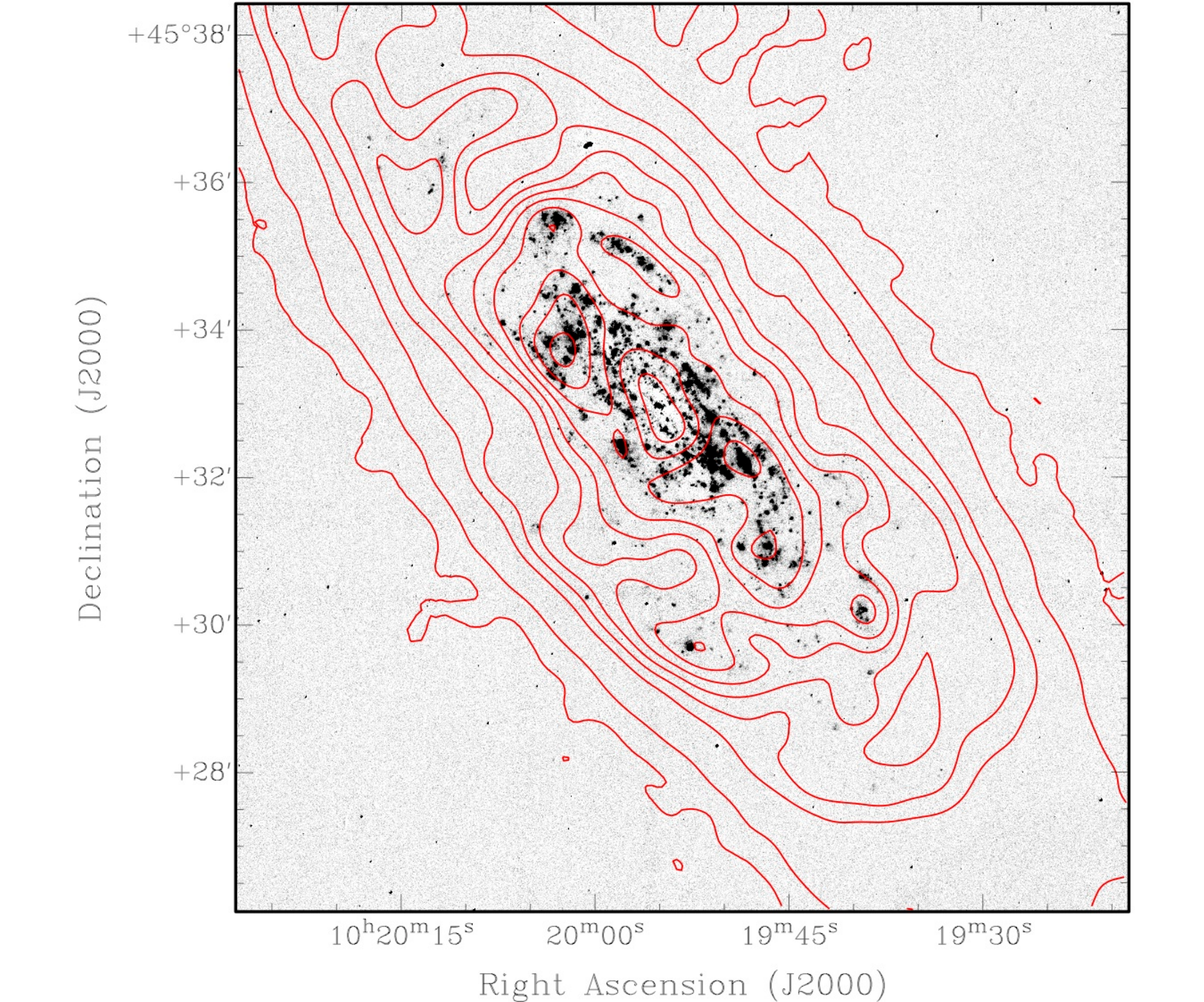}
\end{center}
\caption{
H$\alpha$ image overlaid with contours of the total H{\sc i} map. The
contour levels are chosen to enhance the central parts of the H{\sc i} disk: 
1$\times$10$^{19}$, then (1, 5, 7.5, 10, 12.5, 15, 17.5, 20, 22.5, 25) $\times$10$^{20}$.  
}
\label{halpha_mom0}
\end{figure}

\begin{figure}
\begin{center}
\includegraphics[width=0.49\textwidth]{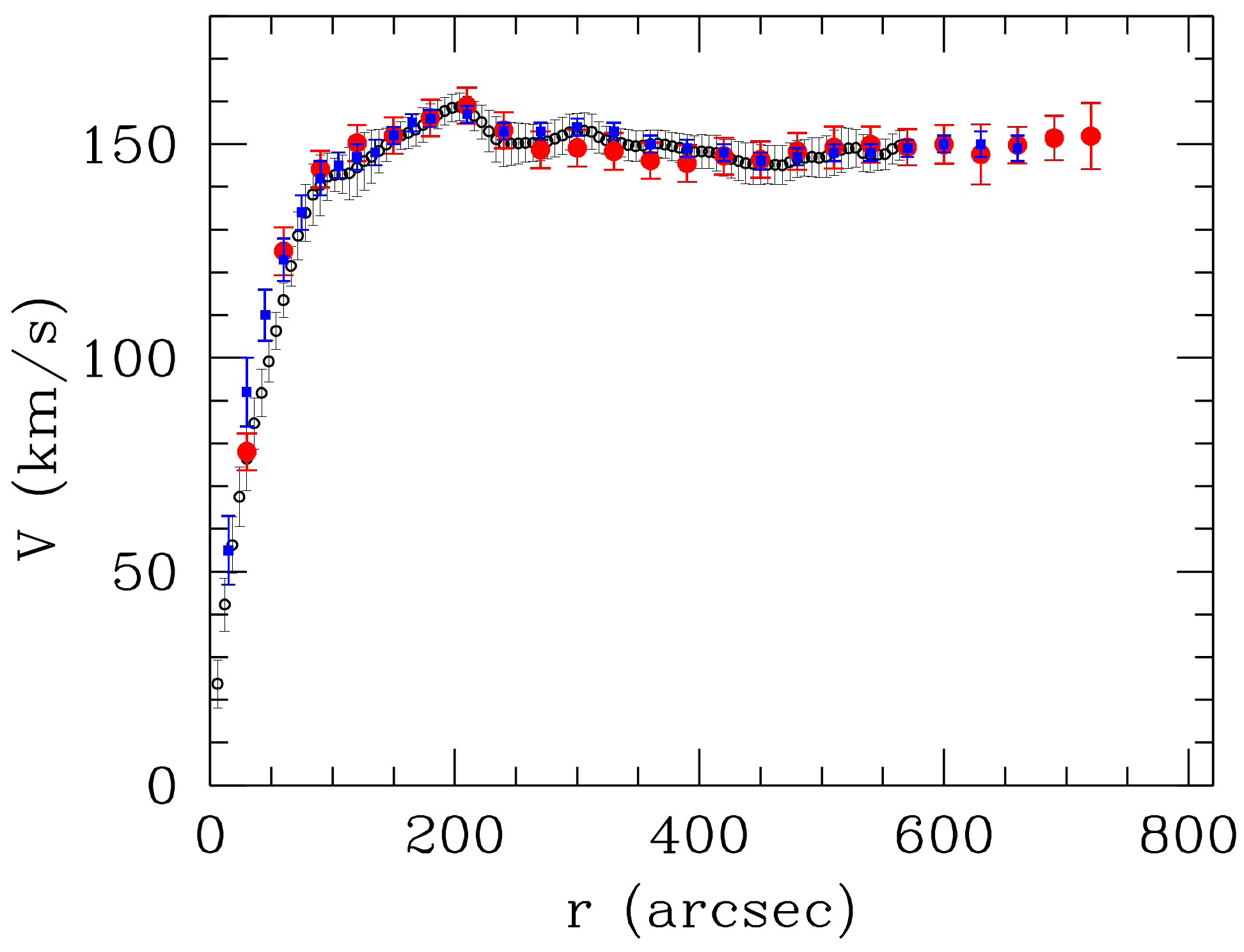}
\end{center}
\caption{
Rotation curve derived here (full red points) compared with the
rotation curve derived by de Blok et al. (2008) from the THINGS data
(open black points) and by Begeman (1989) using less deep WSRT data
(blue squares). 
}
\label{rc_comparison}
\end{figure}

\section{Models}
\label{sect_models}

To investigate in detail the morphology and kinematics of the H{\sc i} in
this galaxy, we resorted to making models of the whole data cube: it
is the most thorough approach that can distentangle subtle differences
between various representations of the H{\sc i} disk. The uncertainties on
the derived parameters are discussed in Section \ref{sect_uncertainties}.

The first model we made was based on the parameters derived by de Blok
et al. (2008): as can be seen in Fig. \ref{channels_data_things}, the
general features of the data cube are reproduced, but these parameters
cannot account for the detailed structure of the emission in each
channel map, which is partly expected as in the present paper we
detect more diffuse emission which typically has different
kinematics. 

To improve the model data cube and obtain a better match to the data,
we decided to follow the strategy described below, based on successive
approximations.  

First, we derived a rotation curve based on a tilted-ring modelling
on the intensity-weighted velocity field (Begeman 1989), leaving the
position angle, the inclination, the systemic velocity and the
rotation velocity as free parameters for each ring. Each side
(approaching and receding) was treated separately and the best-fit
systemic velocity was found to be 657 km s$^{-1}$. Using the
intensity-weighted velocity field can potentially underestimate the
inner rotation curve, especially if the velocity gradient is high, as
is the case here in the inner regions of NGC 3198. Thus we fixed the
position angle and inclination from this fit, and fitted the rotation
curve on the velocity field derived using the WAMET method (Gentile et
al. 2004), a modified version of the envelope tracing method. For
galaxies with a high inclination (such as NGC 3198) or with a poor
resolution it is a more reliable choice than the more traditional
intensity-weighted mean, which in NGC 3198 might give velocities that
are slightly 
biased
towards the systemic velocity. 
The orientation parameters from the intensity-weighted
velocity field were also used to derive the H{\sc i} surface density
profile by averaging the total H{\sc i} map over ellipses. Concerning the
gas distribution in the vertical direction, we used a
sech$^{2}$ distribution, initially with a scale height of 0.2 kpc.

Then, these parameters were used as an initial guess to make 
a fit of the whole data cube using the TiRiFiC (Tilted Ring Fitting
Code) software
(J\'ozsa et al. 2007), leaving the inclination, the position angle and
the rotation velocity free for each ring, separately for each side,
and the velocity dispersion (including instrumental effects) free as a
global parameter. The best-fit velocity dispersion found by TiRiFiC is
11.7 km s$^{-1}$. 

Because of the very high number of degrees of freedom, the parameters
of the best-fit model data cube from TiRiFiC tend to have some strong
discontinuities and unphysical ``jumps'' as a function of radius. Also,
by construction TiRiFiC uses a $\chi^2$ minimisation, which naturally (for
a constant noise) gives a higher weight to high signal-to-noise
regions. However, these regions are not necessarily the most
discriminant between models, that is why the TiRiFiC fitting process
has to somehow be manually guided. 

Therefore we used this set of parameters as a basis to adjust the
parameters manually when necessary, in order to obtain at the same
time a good match of the data cube and a realistic set of
parameters. The inclination and the position angle were adjusted based
on the total H{\sc i} map, and the rotation curve was adjusted based on the
position-velocity diagram along the major axis. Changes of at most 5
km s$^{-1}$ were made to obtain at better match to the observations. 

The emission in some channels is skewed (channels 525.9 -- 591.8 km
s$^{-1}$ and 707.2 -- 740.2 km s$^{-1}$, see Fig. \ref{channels}),
without an apparent change in the 
global position angle, and with no sign of position angle change at
other azimuths at these radii. 

Spekkens \&
Sellwood (2007) 
introduced a formalism to fit velocity fields, focussing on
bisymmetric distortions, which they deem as being more realistic than
the more commonly used radial flows. The reason is that while these
two models give similarly good results, they consider radial flows to
be less
physically plausible as they would have no clear physical origin, and
because of the epicyclic approximation they are restricted to small
non-circular motions. Therefore, we interpret the skewness of the
emission in channels 525.9 -- 591.8 km s$^{-1}$ and 707.2 -- 740.2 km s$^{-1}$ as
a bisymmetric distortion of the velocity field, possibly due to an
elongation of the potential, ignoring the radial term that is not
needed by the present data. In such a model the velocity V$_{\rm model}$ at a
given position in the galaxy disk is modelled as:  

\begin{equation}
V_{\rm model}= V_{\rm sys} + {\rm sin}~i~ [V_{\rm rot} {\rm ~cos}~\theta
- V_{\rm 2,rot} {\rm ~cos}~(2 \theta_b) {\rm ~cos~}\theta]
\label{eq_bisymmetric}
\end{equation}

where $V_{\rm sys}$ is the systemic velocity, $V_{\rm rot}$ is the
rotation velocity, $V_{\rm 2,rot}$ is the $m=2$ additional term,
$\theta$ is the azimuthal angle relative to the projected major axis,
and if $\phi_b$ is the angle between
the bisymmetric distortion and the projected major axis, then 
$\theta_b = \theta-\phi_b$. Such a distortion, when included in the
model, indeed contributes to 
improving the match between the model and 
the observations. It turns out that the amplitude $V_{\rm 2,rot}$ must
be $\sim$15 km s$^{-1}$ beyond a radius of approximately 510 
arcsec, and $\phi_b$ must be $\sim$45$^{\circ}$ to avoid having skewed emission in the
central channels (around the channel at 657.8 km s$^{-1}$) but to have it
at channels 525.9 -- 591.8 km
s$^{-1}$ and 707.2 -- 740.2 km s$^{-1}$. A model with radial motions
(limited to a wedge around the major axis, to avoid having skewed
emission in the central channels) gives almost identical results to this
bisymmetric distortion.

A position-velocity diagram along the major axis with this model
is shown in the left panel of Fig. \ref{pvds}. 
As clearly visible from the position-velocity diagram, to reproduce
the observations it is necessary to add some lagging gas. The first
attempt was a single thick disk with with a scale-height of 1 kpc
and a lag of 18 km s$^{-1}$ kpc$^{-1}$ in the approaching side and 8 km s$^{-1}$
kpc$^{-1}$ in the receding side (Fig. \ref{pvds}, middle panel). This model
reproduces the observations better than the single thin disk, but it
is not satisfactory: by increasing the rotation curve in the inner
parts (to match the outermost contours), there would be lagging gas
missing; by increasing the amplitude of the lag the outer regions
would not be reproduced correctly.

\begin{figure*}
\centering
\begin{minipage}[b]{0.45\textwidth}
\includegraphics[width=\textwidth]{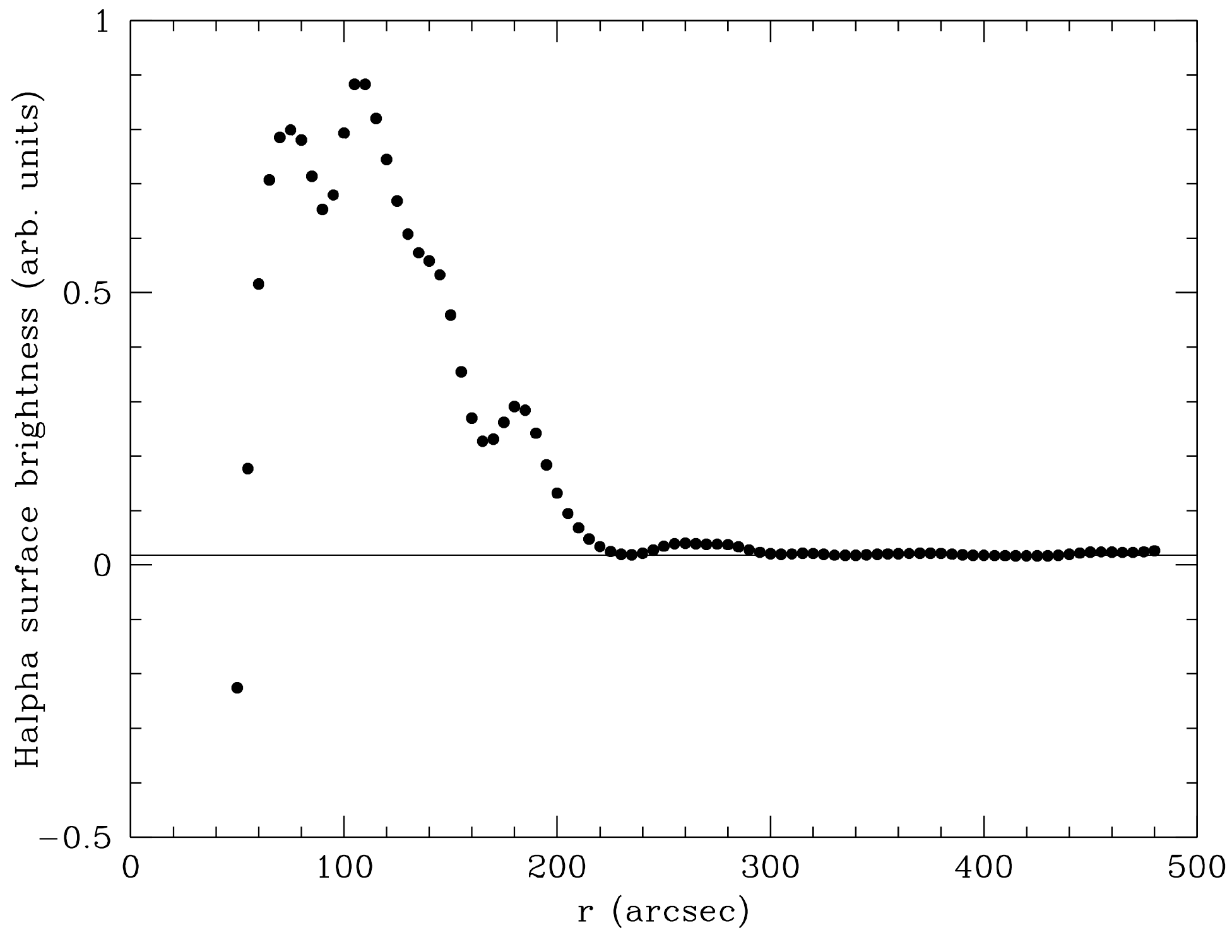}
\end{minipage}
\hspace{2em}
\begin{minipage}[b]{0.45\textwidth}
\includegraphics[width=\textwidth]{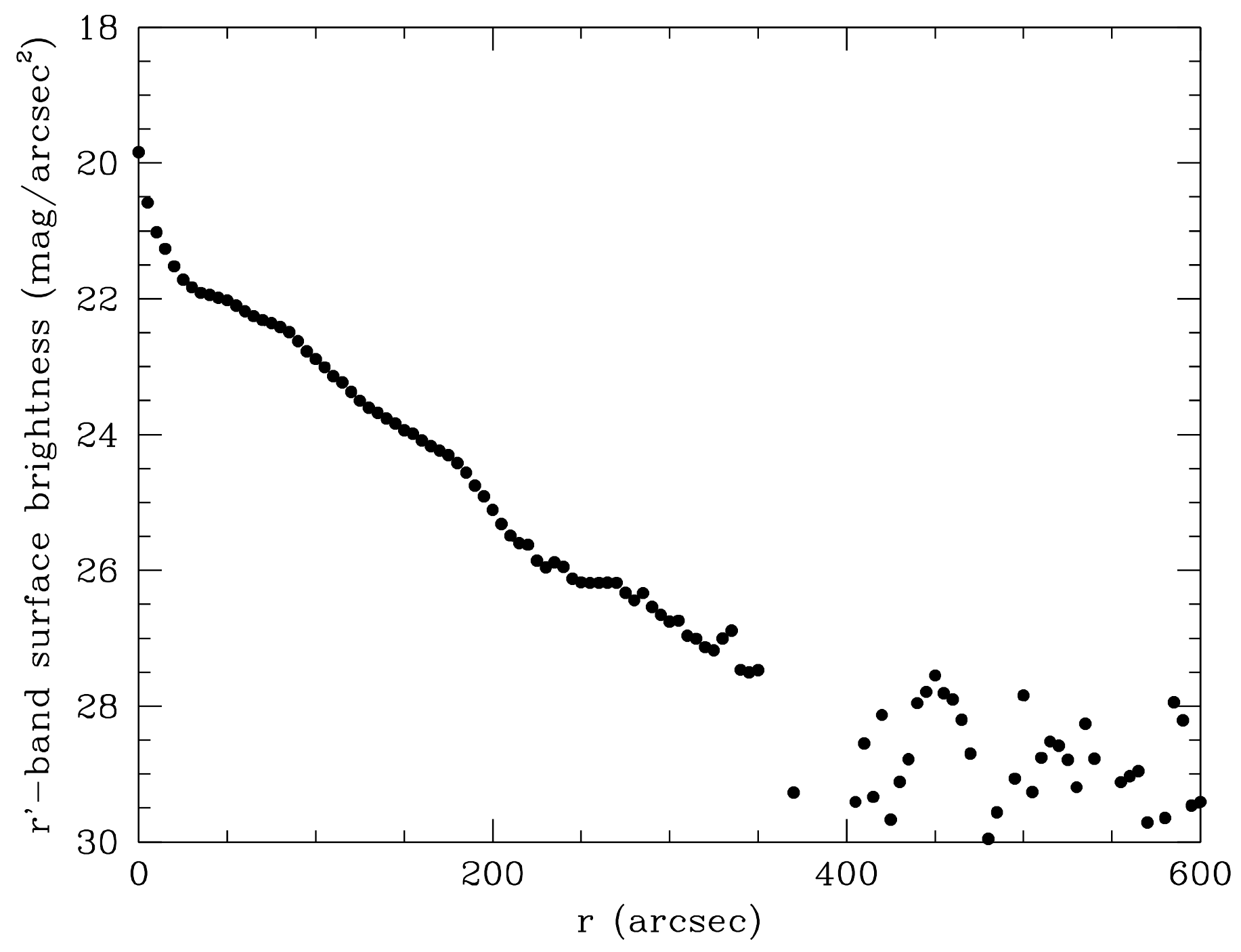}
\end{minipage}
\caption{Azimuthally-averaged surface brightness profiles of the H$\alpha$
  image (left) and our HALOSTARS r'-band image (right). }
\label{iscales}
\end{figure*}

Therefore we resorted to a two-disk model, with a a thick disk that
rotates more slowly than the main (thin) disk, with the same
orientation parameters and run of surface density. This is shown in
the right panel Fig. \ref{pvds}, where a model with a lagging thick disk
component with a scale-height of 3 kpc is a better representation of
the observations. The lag in the model was derived to be 15 km s$^{-1}$
kpc$^{-1}$ in the approaching side and 7 km s$^{-1}$ kpc$^{-1}$ in the receding
side. The thick disk contains about 15\% of the total H{\sc i} mass
and its surface density distribution is a scaled down version of
  the total H{\sc i} density distribution. 

The comparison between the final model data cube and the observed one
is shown in Fig. \ref{channels_data_final}. Because the lagging extraplanar gas emission is
only 1--2\% of the peak in each channel map, we now investigate whether
it could be an artifact due to the CLEAN deconvolution and improper
modelling of the dirty beam. To this aim, we build a model data cube
based on our final model data cube, only without the thick disk. We
then convolve it with the dirty beam, we add some Gaussian noise (with
an rms chosen in order to obtain the same rms noise as in the
observations), and we CLEAN and restore this cube. If the lagging
extraplanar gas had been an artifact, in Fig. \ref{pvd_dirtybeam} we would have seen
some ``pseudo extraplanar gas''. However, we do not, therefore we
conclude that the observation of lagging extraplanar gas is genuine
and is not an artifact due to the dirty beam. This is also confirmed
by the hint of a signal from the extraplanar gas that we see in the
high-resolution cube (see Fig. \ref{pv_majax_hires}). 

The best-fit parameters are plotted in Fig. \ref{best_model_param}
(only the parameters with a radial dependence):
the two sides are quite symmetric.

\subsection{Uncertainties on the derived parameters}
\label{sect_uncertainties}

Uncertainties on the parameters that describe the model data cubes
were estimated by varying the parameter in question un- til the model
data cube becomes only marginally consistent with the observed one,
focussing on a particular projection of the data cube or a particular
region that is particularly sensitive to that parameter. 

First, the inclination and position angle are better derived by
comparing total H{\sc i} maps: a
variation in inclination or position angle of only 2$^{\circ}$ makes
the model total H{\sc i} map barely consistent with the observed one
(see Fig. \ref{mom0_comparison_incl} for the inclination). 

In a
similar way, the (global) velocity dispersion is ideally constrained by
position-velocity diagrams. It turns out that the high-velocity
contours of the position-velocity diagram along the major axis become
too spaced out if the velocity dispersion of 11.7 km s$^{-1}$ changes
by more than 2 km s$^{-1}$.

The uncertainties on the rotation curve were assumed to be 
half the difference between the two sides. To avoid having
unrealistically small errors when the velocities of the two sides are
close, we also considered a minimum error of 4.12/sin($i$) km s$^{-1}$
where $i$ is the inclination angle and 4.12 km s$^{-1}$ is the channel
increment of the data cube we used. The amplitude of bisymmetric
distortion can be constrained from the 
channel maps: we find that it is quite uncertain (15$\pm 10$ km
s$^{-1}$).

How robust are the figures about extraplanar gas? 
For
the thin disk we assumed a scale-height of 0.2 kpc, and for the thick
disk we tried different values. For a given inclination (constrained
from the total H{\sc i} map, see above), the central channels are particularly
sensitive to the scale-height, and we find a
value of $3\pm1$ kpc. 
The relative mass of the extraplanar gas has to be in the range
10\%-20\%, and the amplitude of the lag is not strongly constrained
(NGC 3198 is not edge-on): we find an uncertainty of approximately 5
km s$^{-1}$ kpc$^{-1}$. These uncertainties include the degeneracy between
scale-height and lag amplitude: a small scale-height can be partly
compensated by a large lag value and viceversa. However a close
inspection of the data cube helps resolving this ambiguity and the
above uncertainties take it into account. 

There are however two additional uncertainties in the computation of
the lag value. First, the lag amplitude might vary with radius
(e.g. NGC 891, Oosterloo et al. 2007): we cannot exclude a radially
varying lag amplitude with values up to 10 km s$^{-1}$ kpc$^{-1}$ higher in the
inner parts. Second, the value of the lag is also somewhat degenerate
with the velocity dispersion of the thick disk: a higher value of the
latter can be compensated by a lower value of the lag. However, the
velocity dispersion of the thick disk cannot be higher than about 20
km s$^{-1}$ (and correspondingly a lag about 25\% lower than the
abovementioned values), in which case the emission in the central
channels becomes too extended. Lastly, a thick disk with no vertical
gradient in the lag but whose rotation velocity is 20 km s$^{-1}$ than the
thin disk yields a model that is almost as good as the model with a
vertical gradient in the lag.

\section{Discussion}

\subsection{Extraplanar gas}

We find that the presence of anomalous gas is required to match the
observed and model data cubes. The anomalous gas (defined as the
modelled thick disk) has a mass of 15$\pm$5\% of the total H{\sc i} mass, which
is comparable to previous studies that use very deep H{\sc i} observations
to study extraplanar gas, e.g. Zschaechner et al. (2011) and
Fraternali et al. (2002). The lag is also a required feature of the
model, meaning that the extraplanar gas is progressively rotating
more slowly with increasing vertical distance from the mid-plane. The
value of the lag is comparable with previous estimates,
e.g. Oosterloo et al. (2007) and Zschaechener et al. (2011).

\begin{figure}
\begin{center}
\includegraphics[width=0.45\textwidth]{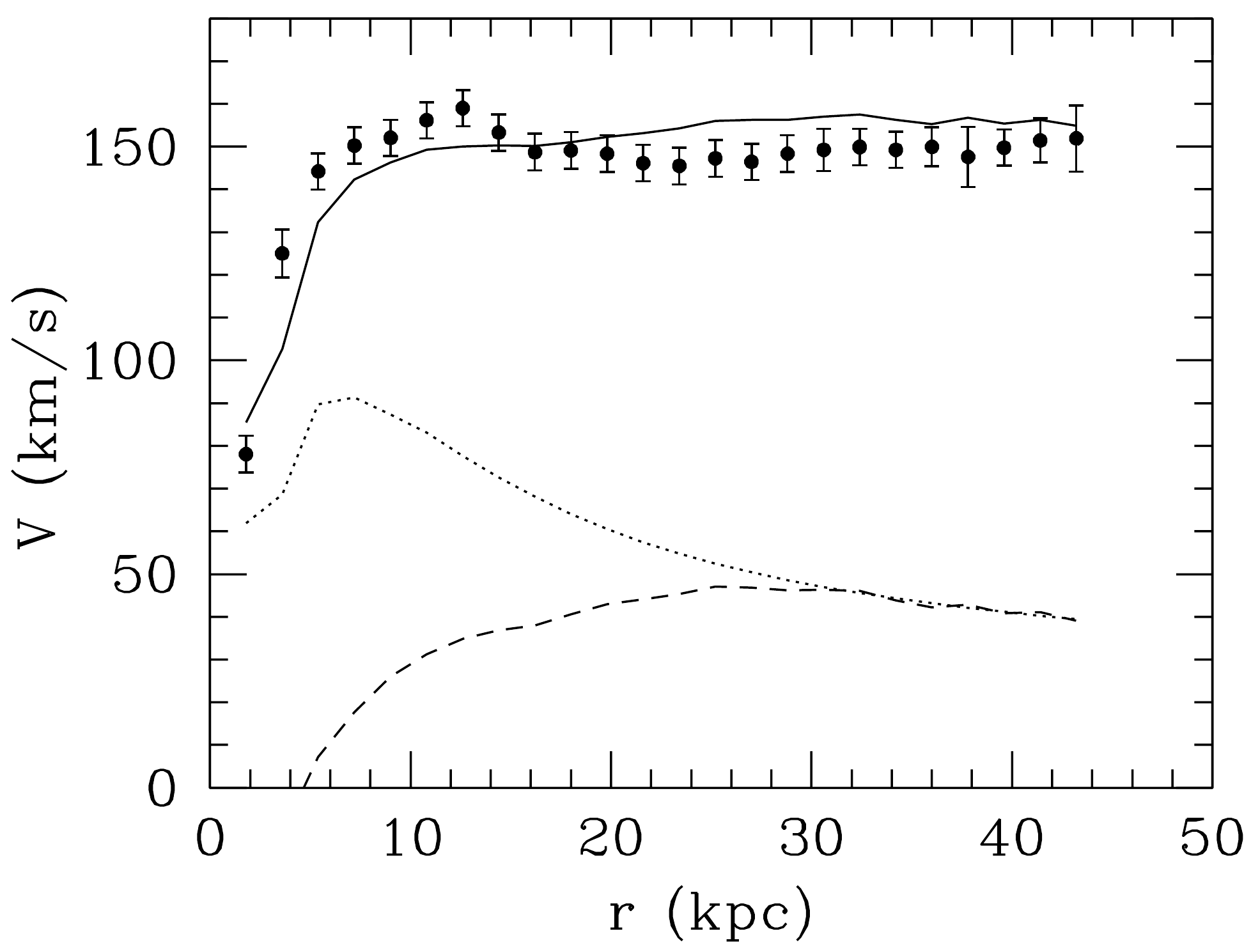}
\end{center}
\caption{
MOND fit of the rotation curve derived in the present paper, using a
distance free within the uncertainties of the Cepheids method
(13.8$\pm$1.5 Mpc). The dashed line is the Newtonian contribution of
the gaseous disks (thin disk and extraplanar gas), the dotted
line is the Newtonian contribution of the stellar disk (from de Blok
et al. 2008) and the solid line is the best MOND fit.
}
\label{rcfit_3198_simple_distconstr_thinthick.pdf}
\end{figure}

The H$\alpha$ image of section \ref{sect_halpha}
is shown in Fig. \ref{halpha_mom0}. There is a
clear correspondence between the brightest H{\sc i} peaks and H$\alpha$
emitting regions. 
How does the radial extent of extraplanar gas compare with optical
properties? HyperLeda gives an $R_{25}$ of 194 arcsec, whereas the RC3
gives 255 arcsec. From the position-velocity diagram along the major
axis (Fig. \ref{pvds}) there is evidence for extraplanar gas out to
5--6 arcmin (receding side) and at least 6--7 arcmin (approaching
side). This can also be compared to the extent of H$\alpha$ emission
(see Fig. \ref{halpha_mom0} and Fig. \ref{iscales}, left panel), and to
our deep r'-band image from the HALOSTARS survey
(Fig. \ref{n3198_mom0_halostars}, Fig. \ref{iscales}, right
panel). Since a precise determination of the extent of the extraplanar
gas is problematic because it is quite model-dependent, an accurate
comparison of the various radial extents (extraplanar gas, stars,
H$\alpha$ emission) is hardly possible in a quantitative manner. From
the surface brightness profiles (Fig. \ref{iscales}) it appears that the H$\alpha$
emission extends out to $\sim 3.5$--5 arcmin (note the central negative
values due to slightly imperfect continuum subtraction, which is
irrelevant for this paper), whereas the r'-band image has some very
faint emission out to $\sim$6 arcmin, comparable to the extent of the
extraplanar gas. 

Therefore, we conclude that the extraplanar gas
extends slightly beyond the actively star-forming body of NGC 3198 but its
extent is comparable to the extent of the stellar disk. This extent
is also seen in other galaxies with detected extraplanar gas, e.g. NGC
2403 (Fraternali et al. 2002) and NGC 6946 (Boomsma et al. 2008), and
it is expected in galactic fountain-type models such as the dynamical
models presented by Fraternali \& Binney (2006): regardless of the
details of the models, during the initial phases of its motion outside
the plane, the gas moves outwards because of the vertical change in
the gravitational potential. The origin of extraplanar gas in galaxies is
still not unambiguously assessed (as explained more thoroughly in
Section \ref{sect_intro}), but its properties in NGC 3198 are 
similar to those derived for other galaxies, e.g. the Milky Way (Marasco \& Fraternali
2011), which were modelled in the framework of galactic fountain-type
models.

\subsection{Rotation curve}

The rotation curve we find (Fig. \ref{rc_comparison}) is roughly consistent
with the one derived by de Blok et al. (2008) and Begeman (1989). 
The former focussed more on the high resolution of their data,
whereas in the present paper the data have a lower angular resolution
but a higher sensitivity to extended emission. In particular, we
confirm the velocity decrease by 10--15 km s$^{-1}$ between 200 and
250 
arcsec, and we find a meaningful rotation curve out to (at least) 720
arcsec. The last useful radius is not an unambiguous quantity to
determine. In order to fully exploit the data, but at the same time to
avoid overinterpreting them, we defined (rather conservatively) the
last reliable radius as the average be- tween the last radius where
the average surface density is above $1 \times 10^{19}$ atoms cm$^{-2}$ (650 arcsec)
and the last radius where the tilted-ring fit on the velocity field
converged. At these large radii (720 arcsec correspond to 48 kpc for a
distance of 13.8 Mpc) there is still no sign of decrease of the
rotation curve. Note that in Begeman (1989) the last few points of the
rotation curve were derived by assuming the same inclination and
position angle as those at a radius of 9.5 arcmin. Moreoever, at these
radii in the receding side, Begeman's fit to the velocity field is
only based on positions outside the major axis.

\subsection{Mass modelling in MOND}

The MOND (Modified Newtonian Dynamics) paradigm was introduced by
Milgrom (1983) as an explanation (alternative to dark matter) for the
absence of Keplerian decline in the observed kinematics of
galaxies. MOND has a remarkable predictive power on galactic scales
(see e.g. Famaey \& McGaugh 2012 and references therein), even though
on larger scales also MOND needs some invisible mass. The galaxy
studied in the present paper, NGC 3198, was claimed to show tension
with MOND (Bottema et al. 2002, Gentile et al. 2011). It is an ideal
case study for MOND because of its inclination (perfectly suited for
kinematical studies, see Begeman 1989), its relative closeness, the
fact that it is a late-type spiral galaxy with regular and symmetric
kinematics, and the accurate determination of its distance using
Cepheids (Kelson et al. 1999, Freedman et al. 2001). Indeed, the MOND
fit is extremely sensitive to the assumed (or fitted) distance. In
MOND, the gravitational acceleration $\vec{g}_{N}$ produced by the visible
matter is linked to the true gravitational acceleration $\vec{g}$ through
the interpolating function $\mu$:  

\begin{equation}
\mu\left(\frac{g}{a_{0}}\right)\vec{g} = \vec{g}_{N},
\label{eq:A}
\end{equation}
where
$\mu(x) \sim x$ for $x \ll 1$ and
$\mu(x) \sim 1$ for $x \gg 1$. Recently preference has been given to
the so-called ``simple'' $\mu$ function: $\mu(x) =\frac{x}{1+x}$.

The contribution of the stellar disk was taken from de Blok et
al. (2008), where is was derived from 3.6 $\mu$m observations
(including a correction for a possible mass-to-light ratio gradient, based on
the J-K colour gradient), 
and the contribution of the
gas from the model data cube discussed above: thin disk
and the thick disk were treated separately (each with its own mass and
scale height), and then their velocities
were added quadratically. 

We analyse here in some more detail the MOND fit with the distance 
left free within the uncertainty of the Cepheids determination.
In comparison with the fit that used the THINGS data (Gentile et
al. 2011, Fig. 5), the present MOND fit is of comparable overall quality:
see Fig. \ref{rcfit_3198_simple_distconstr_thinthick.pdf}. However,
some differences can be seen: in the innermost parts MOND fits the
THINGS rotation curve better than the HALOGAS rotation curve, whereas
in the outer parts the situation is reversed. 
We note that MOND fits each rotation curve best where each rotation
curve is expected to be
the best choice: THINGS in the inner parts (because of the higher
resolution) and HALOGAS in the outer parts (because of the better
sensitivity to the outer, fainter emission).
The best-fit values of the
two fits are the same (best-fit 3.6$\mu$m mass-to-light ratio of 0.48
and best-fit distance of 12.3 Mpc, at the lower end of the allowed
range). The two fits differ in the shape of the contribution of the
gas (apart from, obviously, the rotation curve itself): 
with the more extended data from HALOGAS, we account better for the
outer surface density profile of the gas, and therefore we trace better the
region where the contribution of the gas start declining. Also, none
of the two fits manage to reproduce the decrease in rotation velocity
between 200 and 250 arcsec, which corresponds to the end
of the brighter part optical disk.

\section{Conclusions}

We have presented new, very deep (10 $\times$ 12 hours) H{\sc i} observations of
the spiral galaxy NGC 3198. The observations are part of the HALOGAS
(Westerbork Hydrogen Accretion in LOcal GAlaxieS) survey, see Heald et
al. (2011). These new observations go significantly deeper than
previous H{\sc i} observations. 

We made careful 3D models
of the H{\sc i} layer in NGC 3198, including not only traditional features such as a
variable inclination, position angle, and the rotation curve, but also newer features such as
variable rotation speed as a function of distance from the plane, a
thicker disk and a bisymmetric distorsion of the kinematics. In this
manner we managed to obtain a model that matched the observed data cube very well.

We revealed for the first time in this galaxy the presence of
extraplanar gas over a thickness of a few ($\sim 3$) kpc. Its amount is approximately
15\% of the total mass, 
and one of its main properties is that it appears to be rotating more
slowly than the gas close to midplane, with a (rather uncertain) rotation velocity
gradient in the vertical direction (lag) of 7--15 km s$^{-1}$ kpc$^{-1}$.

Despite the uncertainty on its actual radial extent, the extraplanar
gas seems to be slighly more extended than the star-forming part of
the galaxy (as revealed by the H$\alpha$ image), but our new deep r'-band
image reveals that there is stellar emission out to the end of the
detected extraplanar gas. We detect thin disk H{\sc i} out to a radial
extent that is twice as far as the stellar disk and extraplanar
layer. We also detect a faint H{\sc i} complex beyond the H{\sc i} disk on the
northern side of NGC 3198, with an estimated mass of $\sim$ 5$\times
10^6$ M$_{\odot}$. 

Finally, we make a mass model in the context of MOND (Modified
Newtonian Dynamics) of the newly-derived rotation curve, which is more
extended than previous determinations. The fit quality is modest,
similarly to previous studies, but the outer parts are explained in a
satisfactory way.

\section*{Acknowledgements}

GG is a postdoctoral researcher of the FWO-Vlaanderen (Belgium). PS is
a NWO/Veni fellow. RAMW and MP acknowledge support for this project
from the National Science Foundation under grant AST-0908126. Based on
observations with the Kitt Peak National Observatory, National Optical
Astronomy Observatory, which is operated by the Association of
Universities for Research in Astronomy (AURA) under co- operative
agreement with the National Science Foundation. 
This publication makes use of data products from the Two Micron All
Sky Survey, which is a joint project of the University of
Massachusetts and the Infrared Processing and Analysis
Center/California Institute of Technology, funded by the National
Aeronautics and Space Administration and the National Science
Foundation. Funding for the SDSS and SDSS- II has been provided by the
Alfred P. Sloan Foundation, the Participating Institutions, the
National Science Foundation, the U.S. Department of Energy, the
National Aeronautics and Space Administration, the Japanese
Monbukagakusho, the Max Planck Society, and the Higher Education
Funding Council for England. The SDSS Web Site is
http://www.sdss.org/. Finally, we thank Renzo Sancisi for useful
comments on an early version of this manuscript.

\end{document}